\documentclass[11pt,letterpaper]{article}
\usepackage{graphicx}
\usepackage{color}
\usepackage{xcolor}
\usepackage{subfig}
\usepackage[width=8.50in, height=11.00in, left=1.00in, right=1.00in, top=1.00in, bottom=1.00in]{geometry}
\usepackage[T1]{fontenc}
\usepackage[utf8]{inputenc}
\usepackage{authblk}

\title{The motion of two identical masses connected by an ideal string symmetrically placed over a corner}

\author[1]{Constantin Rasinariu\thanks{crasinariu@colum.edu}}
\author[2]{Asim Gangopadhyaya\thanks{agangop@luc.edu}}
\affil[1]{{\small Department of Science and Mathematics, Columbia College Chicago, Chicago, IL 60605, U.S.A}}
\affil[2]{{\small Department of Physics, Loyola University Chicago, Chicago, IL 60660, U.S.A}}

\begin{document}
	\maketitle

\begin{abstract}
We introduce a novel, two-mass system that slides up an inclined plane while its center of mass moves down. The system consists of two identical masses connected by an ideal string symmetrically placed over a corner-shaped support. This system is similar to a double-cone that rolls up an inclined set of V-shaped rails. We find the double-cone's motion easy to demonstrate but difficult to analyze.   Our example here is more straightforward to follow, and the experimental observations are in good agreement with the theoretical predictions.
\end{abstract}

\maketitle

\section{Introduction}
Anyone who has tried to hang a coat on a corner of a table knows the futility of such an endeavor. In fact, for a frictionless rectangular table the coat will slide toward the corner even if that corner is tilted upward. This situation is reminiscent of a double-cone that rolls up an incline, which is a nice classroom demonstration. The double-cone moves to lower its center of gravity even while its end points ascend on V-shaped rails \cite{Gallitto-2011}. Due to this seemingly paradoxical motion, a double-cone has significant pedagogical appeal, yet there have been very few quantitative studies of the dynamics of this demonstration \cite{Gallitto-2011,Thomas-2006,Balta-2002,Gandhi-2005}.  This lack of analysis is likely due to the difficulty in identifying the points of contact between the double-cone and the rails \cite{Gandhi-2005}.  Here we provide a simple mechanical model to emulate the motion of the coat on the corner of a table. This example consists of a massless string that connects two hanging masses and slides up an inclined plane. Similar to the motion of the double cone, here, as the string moves up the incline, the center of gravity of the system moves down. However, unlike the double-cone, this system is easy to investigate and is well within the capabilities of undergraduate students.

\section{The system}

\begin{figure}[h!]
	\centering
	\includegraphics[width=8.5cm]{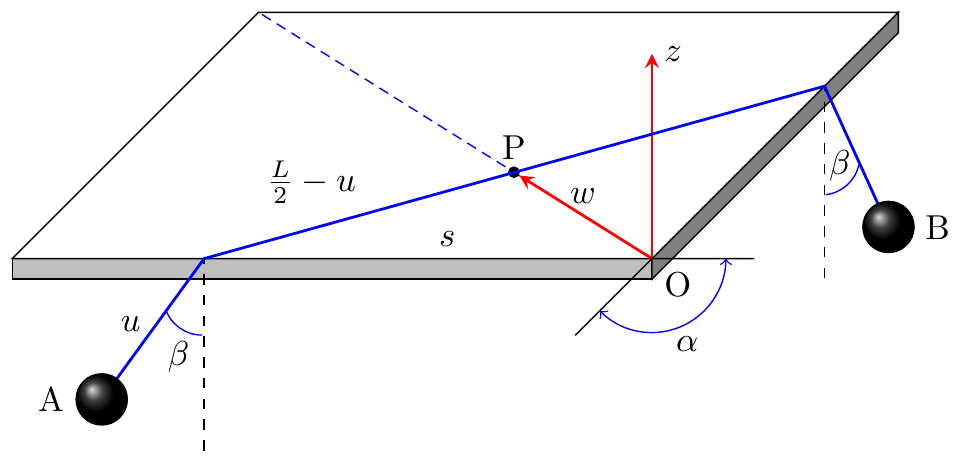}
	\caption{Two identical masses  $A$ and $B$ connected by an ideal string of length $L$, falling symmetrically over the corner of a horizontal table.  Note that point $P$ moves toward point $O$, opposite to the directed segment $\protect\overrightarrow{OP}$.}
	\label{fig:twoBalls}
\end{figure}

We consider two equal masses $A$ and $B$ connected via an ideal string of length $L$ that is symmetrically placed over the corner $O$ of a frictionless table, as depicted in Fig.~\ref{fig:twoBalls}. 

\noindent
The center point of the string is labelled $P$.  Initially the system is at rest with the masses at table height and horizontally separated by the length $L$ of the string. We first investigate the motion of the system assuming a horizontal table. Then we will show that the corner of the table can be tilted upwards by an angle $\theta$ and the point $P$ will still accelerate upward toward the table's corner, and we determine an upper limit for the angle $\theta$.  As the system is released and point $P$ advances toward the corner, we will be interested in finding the acceleration of $P$ as a function of time and as a function of the angle $\alpha$ subtended by the edges of the table at the corner $O$ (see Fig.~\ref{fig:twoBalls}). For this we want to find the motion of the two masses and of their center of mass.

The central point $P$ moves along the bisector of angle $\alpha$, which we take to be the $w$-axis with the origin at $O$ (Fig.~\ref{fig:twoBalls}). The center of mass of the system moves along the $w$-axis towards $O$ and also moves down along the $z$-axis. Initially, the center of mass of the system is  at the level of the table, at $z=0$.  Some time after release, as depicted in Fig.~\ref{fig:twoBalls}, masses $A$ and $B$ have descended by a vertical distance $u \cos \beta$ as point $P$ moves along $w$ towards the vertex $O$.  Here $\beta$ is the angle that the hanging masses make with the vertical, which is indicative of the acceleration of the masses (it will be shown later that this angle is constant). Mass $A$ moves along side $s$ of the table and vertically along $z$, with mass $B$ moving symmetrically along the other edge of the table.  We begin by determining the angle $\beta$ for the general case of a tilted table.

\section{Dynamical Constraints} \label{sec:dynamic}

\begin{figure}[h!]
	\centering
	\includegraphics[width=8.5cm]{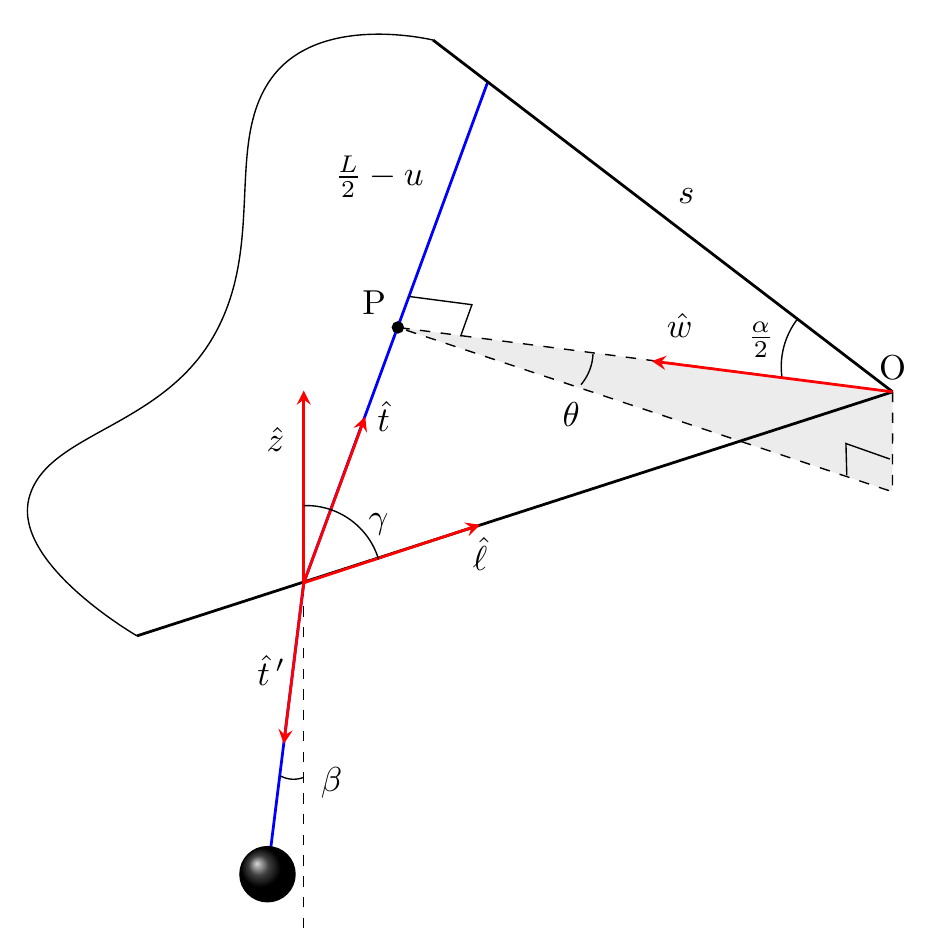}
	\caption{Illustration of the dynamical constraints in the general case of the table tilted by an angle $\theta$. }
	\label{fig:tiltedPlanes}
\end{figure}

We now consider a table that is tilted by angle $\theta$ symmetrically about the bisector of angle $\alpha$ so that corner $O$ is the highest point, as shown in Fig.~\ref{fig:tiltedPlanes}.  Moreover, we assume the string is oriented so that the portion in contact with the top of the table is always horizontal. Because the table is assumed frictionless, the interaction between a piece of string and the table will always be perpendicular to the surface of the table, and we denote by $\vec{N}$ the force of the table on the small portion of the string that is in contact with the edge of the table. 
The tension forces acting on this portion of string are given by $T\,\hat t$ for the segment of the string on the table, and by $T\,\hat t\,'$ for the hanging segment of the string, where $\hat t$ and $\hat t\,'$ are unit vectors that point along these string segments (see Fig.~\ref{fig:tiltedPlanes}).  Note that, owing to our assumptions of a massless string and the smoothness of the table, the tension can be safely assumed to be uniform along the string. Because the string is strictly massless, the net force on any segment must be zero; applying this to the portion of the string in contact with the edge of the table then gives $\vec { N}+T\,\hat t+T\,\hat t\,'=0$.  This equation tells us that $\hat t+\hat t\,' = - \vec {N}/T$, and this vector must be perpendicular to the unit vector $\hat{\ell}$ that points along the edge of the table.

Considering that the motion of ball $A$ is confined  to a vertical plane containing the edge of the table, we see from Fig.~\ref{fig:tiltedPlanes} that we can write $\hat t\,'$ as a linear combination of unit vectors $\hat \ell$ and $\hat{z} $:
\begin{equation}
\hat t\,' = a \,\hat{\ell}+ b\,\hat{z},
\label{eq:ExpansionOf_ell}
\end{equation}
where $a$ and $b$ are constant coefficients. We need to determine these coefficients in order to find the angle $\beta$. Since 
$\hat t\,' $ is a unit vector, we have 
\begin{equation}
	\hat{t}\,' \cdot\hat{t}\,'  = 1 = a^2 + b^2 +2 ab \cos\gamma,
	\label{eq:Coefffcients1}
\end{equation}
where $\cos\gamma = \hat \ell\cdot \hat{z}$.
Moreover, because $\hat t \cdot \hat \ell =\sin\left(\alpha/2\right)$, the fact that $\hat t+\hat t\,'$ is perpendicular to $\hat \ell$ implies that $\hat t\,' \cdot \hat \ell = -\hat t \cdot \hat \ell = -\sin\left(\alpha/2\right)$.  Hence, taking the dot product of $\hat{\ell}$ with Eq.~(\ref{eq:ExpansionOf_ell}), we find that
\begin{eqnarray}
	a+b\, \cos\gamma&=& 	-\sin\left({\frac\alpha 2}\right).
	\label{eq:Coefffcients2}
\end{eqnarray}
Solving Eqs.~(\ref{eq:Coefffcients1}) and (\ref{eq:Coefffcients2}) for $b$ we find	
\begin{equation}
	{\,b\, = \pm \frac{\cos\left(\alpha/2\right)}{\sin\gamma}}
	\label{eq:Coefffcients3}
\end{equation}
and, guided by Fig.~\ref{fig:tiltedPlanes}, we choose the negative sign.
Having found $b$, the coefficient $a$ is then found to be
\begin{equation}
a = -b\cos\gamma -\sin\left({\frac\alpha 2}\right)
=	\frac{\cos\left(\alpha/2+\gamma\right)}{\sin\gamma}.
\end{equation}

We can now determine the angle $\beta$ defined by $\cos\beta = \hat t\,' \cdot (-\hat{z})$. This angle is a crucial parameter that is related to the acceleration in the forward direction because the masses only accelerate horizontally when $\beta \ne 0$. Using Eq.~(\ref{eq:ExpansionOf_ell}), we have
\begin{eqnarray}
\cos\beta = \hat t\,' \cdot (-\hat{z}) &=&-a \cos\gamma - b \nonumber\\
&=&
-\frac{\cos\gamma}{\sin\gamma}\cos\left(\frac\alpha2+\gamma\right)
+ \frac{\cos\left(\alpha/2\right)}{\sin\gamma}\nonumber\\
&=& \sin \left(\frac\alpha2+\gamma\right) \nonumber \\
&=&\cos \left(\frac\pi 2 - \frac\alpha2-\gamma\right), \label{e:Beta0}
\end{eqnarray}

\begin{figure}[h!]
	\centering
	\subfloat[{}
	\label{subfig-1:dummy}]{%
		\includegraphics[width=0.4\textwidth]{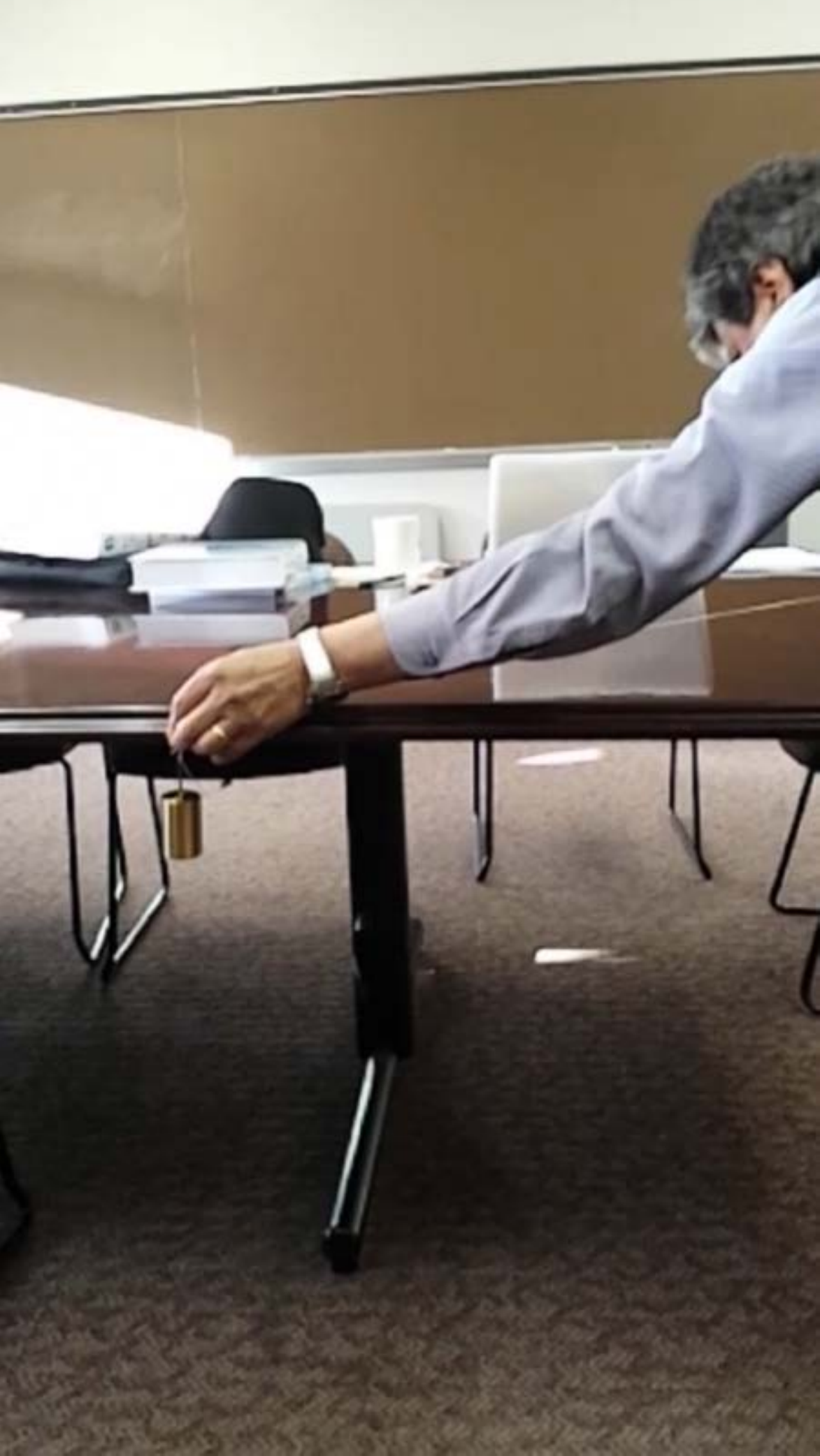}
	}
	\hfill
	\subfloat[{}
	\label{subfig-2:dummy}]{%
		\includegraphics[width=0.4\textwidth]{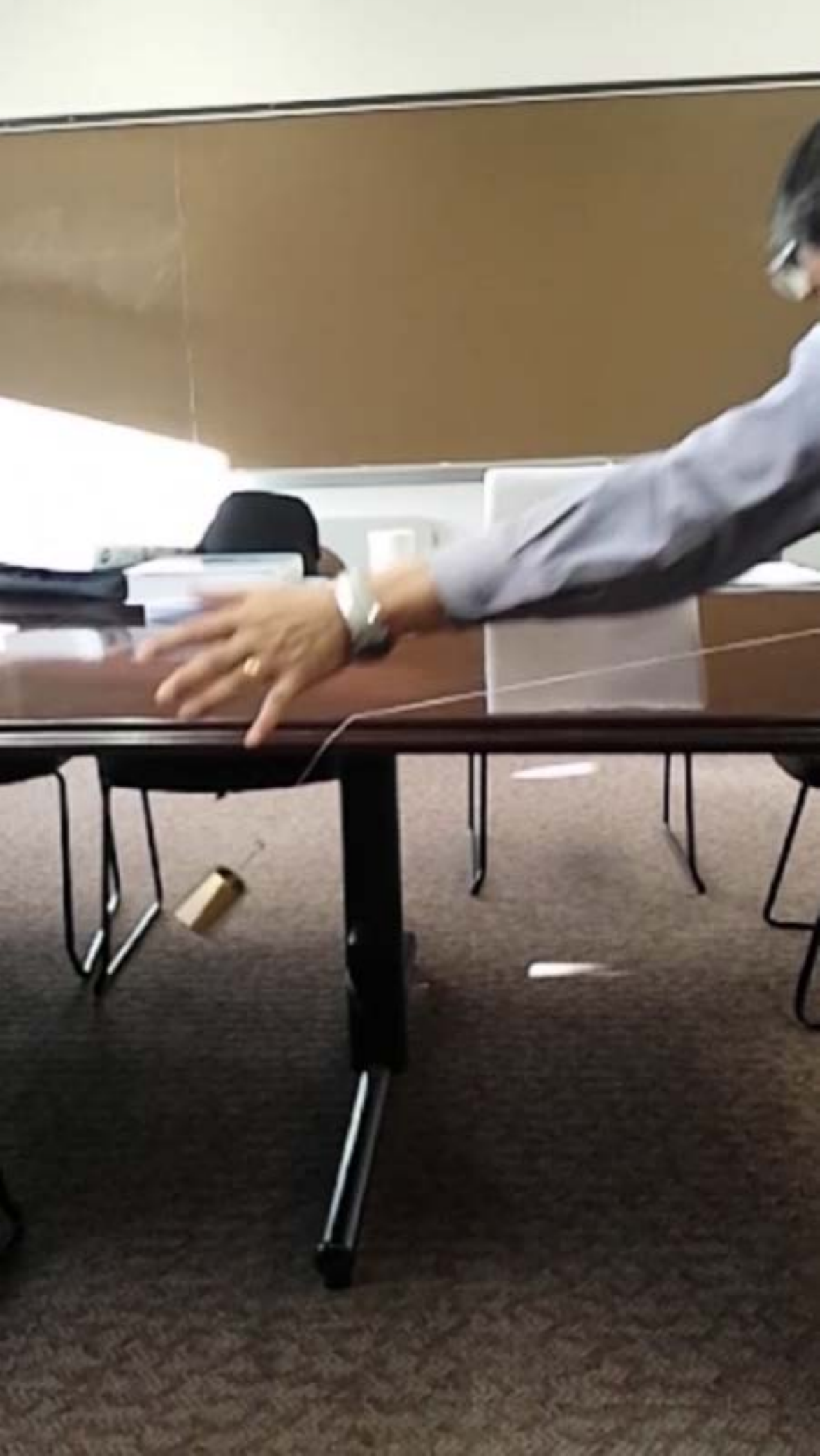}
	}
	\caption{Experimental check: (a) prior to release; (b) just after release.  The angle of the moving string with the vertical is observed to be approximately $\beta=\pi/4$, in agreement with the prediction of Eq.~(\ref{eq:Angle_of_Inclination}), given a table corner angle of $\alpha = \pi/2$. In reality, given the finite size of the mass A, we cannot start from strictly $u=0$.  This generates an oscillation of very small angular amplitude around $\beta=\pi/4$.  }
	\label{fig:Experiment}
\end{figure}
\noindent
which gives
\begin{eqnarray}
\beta  = \pm \left(\frac\pi 2 - \frac\alpha2-\gamma\right)~. \label{e:Beta1}
\end{eqnarray}
Note that for a horizontal table, with $\theta=0$ and $\gamma=\pi/2$, angle $\beta$ should be positive.
Thus, choosing the negative sign in Eq.~(\ref{e:Beta1}) we obtain 
\begin{eqnarray}\beta  =  \frac\alpha2+\gamma-\frac\pi 2~,
\label{e:GammaBeta}
\end{eqnarray} 
which, for a horizontal table, gives
\begin{equation}
\beta =\frac{\alpha}{2} .
\label{eq:Angle_of_Inclination}
\end{equation}

As illustrated in Fig.~\ref{fig:Experiment}, this last relationship can be verified experimentally.  
Note that if we start the system from a different initial configuration where $\beta$ differs from ${\alpha}/{2}$, and the two masses hang symmetrically at a given depth below the table, then the angle $\beta$ will not remain constant, as the onset of the motion will induce oscillations of the moving masses.

\section{The Equations of Motion for a Horizontal Table}

In this section we return the case of a horizontal table, which has $\theta =0 $, $\gamma=\pi/2$, and $\beta = \alpha/2$. From the geometry of the system (see Fig.~\ref{fig:twoBalls}), we have
\begin{equation}
	\label{e:balls1}
	{u} = L/2 - {w}\tan\beta
\end{equation}
and
\begin{equation}
	\label{e:balls2}
	s= {w}\, \sec\beta ~. 
\end{equation}

Mass $A$ moves along side $s$ with speed $\dot{s}$ and along $u$ with speed $\dot{u}$, so its velocity vector is $\vec{v}_A=\dot{s}\,\hat{\ell}+\dot{u}\,\hat{t}\,'$. 
Since the kinetic energy of $B$ is the same as that of $A$, the total kinetic energy of the system is given by 
\begin{eqnarray}
\label{e:balls3}
T&=& 2\left( \frac m2 \right) \left[\dot{s}^2+\dot{u}^2 + 2 |\dot{s}| |\dot{u}| \cos\left( \pi/2 +\beta\right) \right] \nonumber \\
&=& 2\left( \frac m2 \right) \left(\dot{s}^2+\dot{u}^2 - 2 |\dot{s}| |\dot{u}|  \sin \beta \right) \nonumber \\
&=& m \dot{w}^2 \left(
\sec^2\beta + \tan^2\beta-2  \tan^2\beta \right) \nonumber \\
&=& m \dot{w}^2
~.\end{eqnarray}
Meanwhile, taking the surface of the table as the zero point, the potential energy is
\begin{eqnarray}
	\label{e:balls4}
	V&=& -2 m g {u}\cos\beta\nonumber \\
	&=& -2mg\left( L/2-w \tan \beta\right) \cos \beta\nonumber \\
	&=& 2mg w \sin \beta-mg L \cos \beta~.
\end{eqnarray}

Knowing both $T$ and $V$ in terms of the single variable $w$, the Lagrangian ${\cal{L}}=T-V$ for the system is
\begin{equation}
\label{e:balls5}
{\cal{L}}(w,\dot{w}) = m \dot{w}^2 - 2mgw \sin \beta
+ mgL \cos \beta.
\end{equation}
The corresponding Euler-Lagrange equation reads
\begin{equation}
	\label{e:balls6}
	2m \ddot{w}  + 2mg\sin \beta  =0~.
\end{equation}
Thus, the point $P$ moves with a constant acceleration
\begin{equation}
	\label{e:balls_acceleration}
	\ddot{w} = - g  \sin \beta = -g \sin (\alpha/2).
\end{equation}
As one would expect, when $\alpha =0$ the acceleration $\ddot{w}$ vanishes (which explains why we are able to hang our coats on nails), while for $\alpha \to \pi$ it becomes $-g$.  

\subsection{The motion of the masses}
At this point we are well equipped to describe the motion of the two masses. We consider only mass $A$, because the motion of $B$ follows by symmetry.  As the system evolves, from Eqs. (\ref{e:balls1}) and (\ref{e:balls2}) we find that $A$ moves along the edge of the table with  acceleration 
\begin{equation}
	\label{e:balls8a}
	\ddot{s} = \ddot{w}\sec \beta 
\end{equation}
and along $u$ with acceleration
\begin{equation}
\label{e:balls8b}
\ddot{u} = -\ddot{w} \tan \beta.
\end{equation}
To determine the trajectory, we express the position of mass $A$ in terms of the coordinates $\ell$ and $z$, as measured from the origin $O$; that is, the position vector of mass $A$ is given by $\vec{r}_A = \ell(t) \,\hat{\ell} + z(t)\,\hat{z}$.  Assuming the masses are released from rest at the height of the table, we have 
\begin{eqnarray}
\ell(t) &=&  -s_0 - \frac12 \left( \ddot s + \ddot{u}\sin \beta \right)  t^2 \nonumber\\
&=& -s_0 - \frac {\ddot{w}}2 \left( \sec \beta - \tan \beta \sin \beta \right)  t^2
\nonumber\\
&=& - s_0 - \frac{\ddot{w}}{2} (\cos \beta)   t^2 
\nonumber\\
&=& - s_0 + \frac{g}{2}(\sin\beta \cos\beta) t^2,
\label{eq:l(t)}
\end{eqnarray}
where $s_0=(L/2)\,{\rm cosec}\beta$ is the initial position of the mass along the edge of the table, and we have made use of Eq.~\ref{e:balls_acceleration} in the last line.  At the same time, we have
\begin{eqnarray}
z(t) &=& - \frac{1}{2}(\ddot{u}\cos \beta)   t^2 \nonumber\\
&=& - \frac{g}{2}(\sin^2\beta) t^2.
\label{eq:z(t)}
\end{eqnarray}
These equations give a linear trajectory for the mass described by 
\begin{eqnarray}
\ell &=&  -s_0 - z\cot\beta .
\end{eqnarray}

The acceleration of mass $A$ can be found from Eqs.~\ref{eq:l(t)} and \ref{eq:z(t)} as
\begin{eqnarray}
\label{eq:a_A}
\vec{a}_A &=& \ddot{\ell}\,\hat{\ell} + \ddot{z}\,\hat{z} = g\sin\beta\left[ (\cos\beta) \hat{\ell} - (\sin\beta) \hat{z}\right].
\end{eqnarray}
For a horizontal table, the vectors $\hat{\ell}$ and $\hat{z}$ are orthogonal, so the magnitude of the acceleration is $a_A=g\sin\beta=g\sin(\alpha/2)$, which is precisely the acceleration that results when a mass slides down a frictionless incline of angle $\alpha/2$.  Thus, the motion of mass $A$ is quite similar to the motion that would result if the table were folded along the bisector of the corner angle (along $w$ in Fig.~1).  When $\alpha=0$ (and hence $\beta=0$) the system will not move and $a_A=0$.  But as $\alpha \to \pi$, the components $\ddot{\ell} \to 0$ and $\ddot{z} \to -g$ so that $a_A$ approaches $g$ as expected.  Because $B$ moves symmetrically on the other side of the table, the magnitude of its acceleration will be identical: $a_B = a_A$.

Interestingly, from Eq.~\ref{e:balls_acceleration} we see that $a_A = |\ddot{w}|$, which demonstrates that point $P$ also accelerates as if it were sliding down an incline of angle $\alpha/2$. This result is not a simple coincidence. 
\begin{figure}[h!]
	\centering
	\includegraphics[width=12cm]{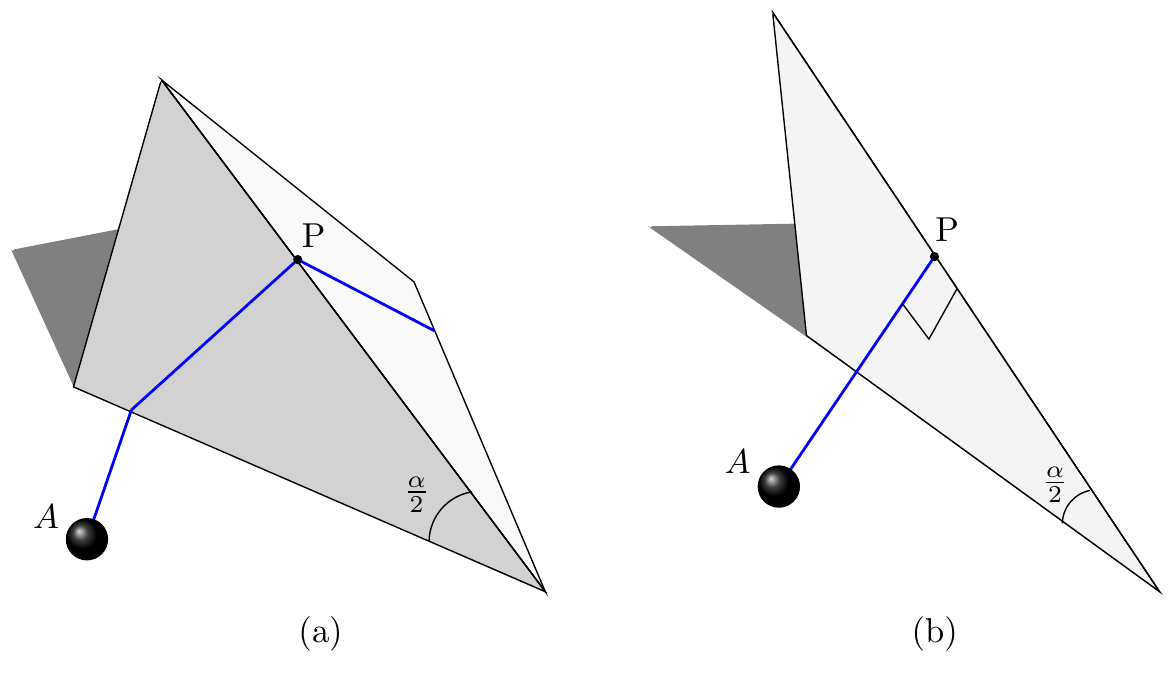}
	\caption{Symmetrically folding up the table along the bisector while keeping its edges horizontal.}
	\label{fig:folding}
\end{figure}
Let us consider that we are folding up the table along the bisector while keeping the edges horizontal as illustrated in Fig. \ref{fig:folding}(a). This (tent-like) structure is completely equivalent to the initial horizontal table, as all the constraints derived in Section \ref{sec:dynamic} remain the same, leading to $\beta = \alpha/2$. Hence, both point $P$ and mass $A$ move with accelerations of the same magnitude $g \sin (\alpha/2)$. In the limiting case of a completely folded table, one simply obtains an incline of angle $\alpha/2$ (see Fig. \ref{fig:folding}(b)). In this case, if the motion is started with the string perpendicular to the incline, it follows that again, both point P and the mass $A$ move synchronously with accelerations of the same magnitude.

\subsection{The motion of the center of mass}

\begin{figure}[h!]
	\centering
	\includegraphics[width=6.5cm]{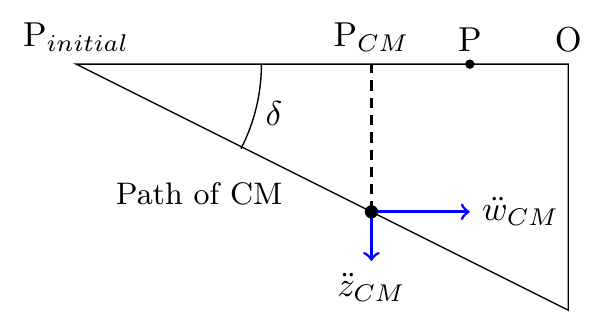}
	\caption{The path of the center of mass of the system below a horizontal table.}
	\label{fig:CM}
\end{figure}

The center of mass of the system moves simultaneously along the $w$ and $z$ axes, as depicted in Fig.~\ref{fig:CM} (see also Fig.~\ref{fig:twoBalls}).  Thus, the center of mass descends on a oblique path making an angle $\delta$ with respect to the horizontal.  

\noindent
To determine $\delta$, we note that mass $A$ hangs at an angle $\beta$ with respect to the vertical axis, as shown in Fig. 2.  If $\vec{r}_A$ and $\vec{r}_B$ are the position vectors for masses $A$ and $B$, respectively, then the center of mass position is given by $\frac{1}{2}(\vec r_A+\vec r_B)$.  As previously discussed, the position of mass $A$ can be written
\begin{equation}
\vec{r}_A = \ell(t) \,\hat{\ell}_A + z(t)\,\hat{z},
\end{equation}
with $\ell(t)$ and $z(t)$ given by Eqs.~\ref{eq:l(t)} and \ref{eq:z(t)}, respectively.  In this equation, we write $\hat{\ell}_A$ to represent the unit vector along the edge of the table for mass $A$.  Similarly, the position of mass $B$ is given by
\begin{equation}
\vec{r}_B = \ell(t) \,\hat{\ell}_B + z(t)\,\hat{z},
\end{equation}
where $\hat{\ell}_B$ is the unit vector along the edge of the table for mass $B$.  Using the geometry of Fig.~2 to note that $\hat{\ell}_A+\hat{\ell}_B=-2\cos(\alpha/2)\,\hat{w}$, the position of the center of mass then becomes
\begin{equation}
\vec r_{CM} = \left[\ell (t)\cos\beta\right] \hat{w} + z(t)\,\hat{z},
\end{equation}
where, for a horizontal table, we have used $\alpha/2=\beta$.

The acceleration of the center of mass can be found by differentiating the position vector twice.  Carrying out the differentiation and again making use of Eqs.~\ref{eq:l(t)} and \ref{eq:z(t)}, we find that the center of mass undergoes a constant acceleration
\begin{eqnarray}
\vec a_{CM} 
&=& \left(\ddot{\ell}\cos\beta\right) \hat{w} + \ddot{z}\,\hat{z} \nonumber \\
&=& -g \sin \beta \left[ (\cos^2 \beta)\hat{w} + (\sin \beta)\hat{z} \right].
\label{eq:a_CM}
\end{eqnarray}
Because the center of mass starts from rest, its trajectory points in the same direction as the (constant) acceleration.  Using the acceleration components in Eq.~\ref{eq:a_CM}, the angle of the center-of-mass trajectory in Fig.~\ref{fig:CM} is found to be
\begin{equation}
\delta=\tan^{-1} \left(\frac{\sin \beta}{\cos^2\beta}\right).
\end{equation}
In the limit $\alpha \to \pi$ (so that $\beta \to \pi/2$), the acceleration of the center of mass becomes $\vec a_{CM} = - g \,\hat{z}$, while the angle of the trajectory approaches $\delta\to\pi/2$, as expected.

\section{Tilted Table}
So far we have been considering the motion of the string on a horizontal plane. We now relax this constraint and find, surprisingly, that the string will slide \emph{up} an inclined plane. For a given corner angle $\alpha$, we can tilt the table upwards at an arbitrary angle $\theta$ by symmetrically raising the point $O$ in such a way so that the two edges of the table intersecting at $O$ make the same angle with the horizontal.  It is then natural to determine the maximum tilt angle $\theta_{\rm max}$ for which point $P$ still moves towards $O$. 

The angle $\beta$ that the hanging mass makes with the vertical (see Fig.~\ref{fig:tiltedPlanes}), expressed in Eq.~(\ref{e:Beta1}) in terms of $\gamma$, is essential for ensuring that there is a forward acceleration (e.g., $\beta=0$ signals that the string is vertical and hence there is no horizontal force on the masses, and hence no pull towards the corner).  Thus, we need to relate $\beta$ to the angle of inclination $\theta$ of the table.  From Fig.~\ref{fig:tiltedPlanes}, because the vectors $\hat \ell$, $\hat t$, and $\hat w$ all lie on the plane of the table, we can deduce that $\hat \ell = \sin (\alpha/2)\,\hat t - \cos (\alpha/2)\,\hat w$. Taking the scalar product of this relation with $\hat{z}$, we get
\begin{eqnarray}
\hat{z}\cdot\hat{\ell} = \cos \gamma = \sin\theta \cos  \left( \frac{\alpha}{2}\right),
\end{eqnarray}
{where we have used $\hat t \cdot \hat{z}=0$ and $\hat w \cdot \hat{z}= - \sin\theta$}.  Meanwhile, from Eq.~(\ref{e:Beta0}) we know that 
\begin{equation}
	\cos\beta 
	= \sin \left(\frac\alpha2+\gamma\right) = \sin \left(\frac\alpha2\right) \cos\gamma+
		\cos \left(\frac\alpha2\right) \sqrt{1-\cos^2\gamma} ~.
\end{equation}
Combining these last two equations, we obtain the connection between the dynamical angle $\beta$, the inclination of the table $\theta$, and the corner angle $\alpha$:
\begin{equation}
	\cos\beta= \sin \left(\frac\alpha2\right) \cos  \left( \frac{\alpha}{2}\right) \,\sin\theta+
	\cos \left(\frac\alpha2\right) \sqrt{1-\cos^2 (\alpha/2)\sin^2\theta}~.
\end{equation}

To find the maximum angle $\theta_{\rm max}$ at which the string will no longer be able to climb, we note that there will be no acceleration towards $O$ when the angle $\beta$ becomes zero, which implies
\begin{equation}
\sin \left(\frac\alpha2\right) \cos  \left( \frac{\alpha}{2}\right) \sin\theta_{\rm max}+
	\cos \left(\frac\alpha2\right) \sqrt{1-\cos^2 (\alpha/2)\sin^2\theta_{\rm max}}= 1~.
\end{equation}
This is a quadratic equation for $\sin\theta_{\rm max}$ that can be solved to give
\begin{eqnarray}
\sin\theta_{\rm max} = \tan \left( \frac{\alpha}{2}\right) \quad \mbox{for} \quad 0\leq \alpha \leq \pi/2 ~.\label{eq:ThetaMax1}
\end{eqnarray}
Thus, for $\alpha = \pi/2$ we find $\theta_{\rm max}=\pi/2$; that is, the string will advance towards the corner for \textit{any} angle of inclination except a vertical table. In a moment we will show that this is also true for $\alpha > \pi/2$.

\begin{figure}[h!]
	\centering
	\includegraphics[width=6.5cm]{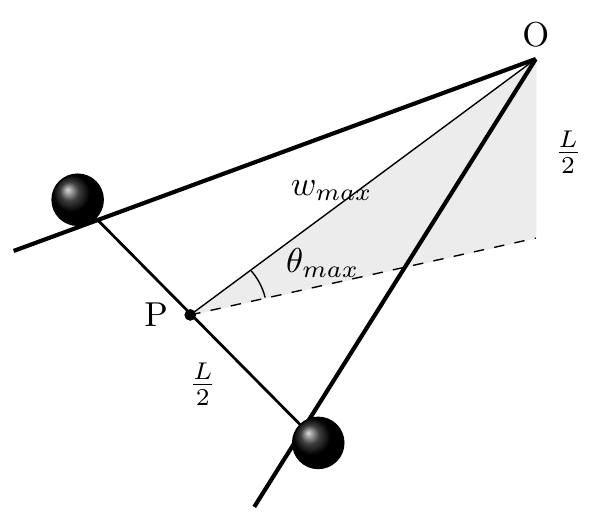}
	\caption{When $\theta=\theta_{\rm max}$, the center of mass moves along the horizontal dashed line.} 
	\label{fig:MaxAngle}
\end{figure}

Note that we could have determined the angle $\theta_{\rm max}$ from the geometry of the table as well. The upward motion of $P$ is possible only if the center of mass of the system moves downward. 
Hence, from Fig.~\ref{fig:MaxAngle} we see that when the inclination $\theta$ reaches the maximum value $\theta_{\rm max}$, the corner of the table should be a distance $L/2$ higher than the initial location of point $P$.  In this geometry, the center of mass will move exactly horizontally as point $P$ moves up the plane, which gives
\begin{equation}
\frac{L}{2}  = w_{\rm max} \sin\theta_{\rm max} = \frac{L}{2} \cot \left( \frac{\alpha}{2}\right) \sin\theta_{\rm max} = \frac{L/2}{\tan(\alpha/2)}\sin\theta_{\rm max} ~.
\end{equation}
Thus, we arrive at the same condition as in Eq.~(\ref{eq:ThetaMax1}).

\begin{figure}[h!]
	\centering
	\includegraphics[width=12.5cm]{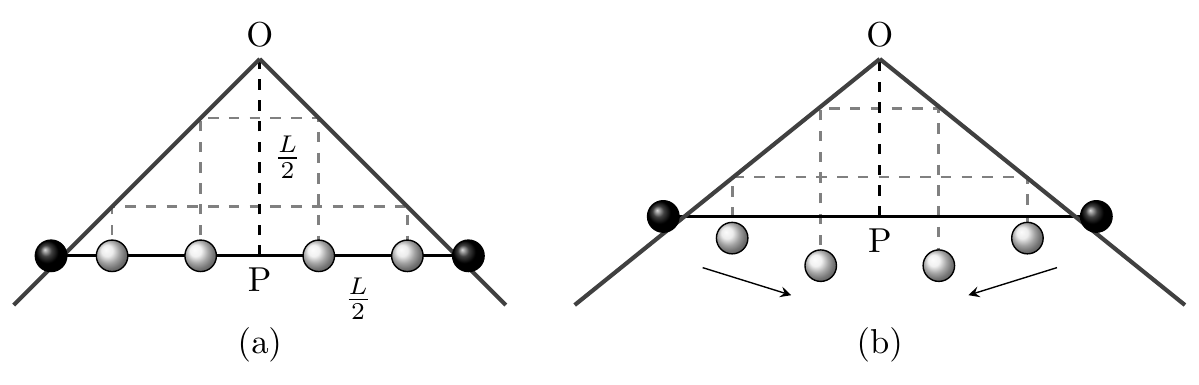}
	\caption{A vertical table, viewed perpendicular to the table top. (a) A table with corner angle $\alpha=\pi/2$ leads to no change in height for the masses. (b) When $\alpha>\pi/2$, the masses, and the center of mass, decrease in height.}
	\label{fig:Vertical}
\end{figure}

Note also that for small values of $\alpha$, we have $\theta_{\rm max}\simeq\alpha/2$, and in general $\theta_{\rm max}>\alpha/2$.  Figure~\ref{fig:Vertical}(a) shows that for $\alpha=\pi/2$ and a vertical table, the masses (and hence the center of mass) will remain at the same height as point $P$ is moved toward $O$. On the other hand, Fig.~\ref{fig:Vertical}(b) shows that for a vertical table with a corner angle $\alpha>\pi/2$ the masses (and hence the center of mass) will \textit{decrease} in height as point $P$ moves toward $O$. In practice, of course, it is impossible to have a completely frictionless apparatus.  Therefore, the motion will most likely cease before reaching a vertical tilt of the table.

\section{Conclusions} 
We analyzed a system that consists of two identical masses connected by an ideal string placed symmetrically over a corner of a frictionless table. On a horizontal table, the string moves towards the corner for any value of the corner angle $\alpha$.  If the table is tilted upward, we find that the string still moves towards the corner provided that the tilting angle is less than a critical value. This system is reminiscent of a double-cone rolling up a set of inclined V-shaped rails. The double-cone's motion, while relatively easy to demonstrate, is rather difficult to analyze. The example considered here is straightforward to understand, and it does not involve the subtleties of the three-dimensional geometry required for the involved analysis of the double-cone problem. Indeed, we find that the corner problem is not without intrigue. If the corner angle is greater than $\pi/2$, then the string will slide up and jump over the corner, even for a  vertical plane.

\section{Acknowledgments} We would like to sincerely thank the anonymous referees for pointing out some mistakes and directing us to experimentation. We owe a debt of gratitude to J\'er\^ome Lodewyck for invaluable comments and to Drag Sampadaka for the careful reading of the text and numerous suggestions. We would also like to thank Thomas Ruubel who patiently helped us in the lab where we verified our theoretical predictions.

\end{document}